\documentclass[12pt,thmsa]{article}
\usepackage{amsfonts}

\input{tcilatex}
\begin{document}

\author{Karl-Georg Schlesinger \qquad \\
Erwin Schr\"{o}dinger Institute for Mathematical Physics\\
Boltzmanngasse 9\\
A-1090 Vienna, Austria\\
e-mail: kgschles@esi.ac.at}
\title{On the universality of string theory}
\date{}
\maketitle

\begin{abstract}
String theory is accused by some of its critics to be a purely abstract
mathematical discipline, having lost the contact to the simple yet deeply
rooted questions which physics provided until the beginning of this century.
We argue that, in contrary, there are indications that string theory might
be linked to a fundamental principle of a quantum computational character.
In addition, the nature of this principle can possibly provide some new
insight into the question of universality of string theory (string theory as
the ``theory of everything'').
\end{abstract}

\section{Universality}

String theory is often also named the ``theory of everything'' since it is
supposed to explain all the fundamental forces and all the elementary
particles of nature, including the gravitational force which - as we know
from Einstein's theory of general relativity - is intimately linked to the
structure of space-time. But could it not be possible that the story goes on
after string theory that we would discover e.g. additional new forces of
nature which it does not explain? After all, even the early 20th century
believed that electromagnetism and gravity are the only fundamental forces,
the discovery of the weak and strong forces came only later since their
presence is basically restricted to the level of elementary particles. This
question is usually answered by refering to the Planck length. The weak and
strong forces were discovered by investigating smaller distance scales than
before but a quantum theory of gravity is supposed to provide a smallest
length scale in the order of the Planck length and string theory is assumed
to be valid down to this scale. Indeed, there are strong arguments that
string theory involves such a minimum length, shorter length scales being
physically non observable in the same sense as the Heisenberg uncertainty
relation shows classical phase space coordinates to be non observable beyond
a certain precision (There is a large body of work on this topic in string
theory. In order to keep the list of references of this short letter
restricted, we refer to \cite{Pol} as a general reference for string theory
and cite only individual research papers on very special topics not included
there.). If we would have a theory which applies down to the Planck scale
and if shorter length scales are physically unobservable, it seems that we
would really have a final physical theory. A critic who is acquainted with
string theory could try to find a loop hole in this argument by pointing out
that string theorists themselves believe that the - yet to be discovered -
fundamental formulation of the theory might not refer to space-time at all.
So, maybe it is simply wrong to assume that gaining new knowledge will
necessarily be linked to the necessity to increase space-time precision of
measurements. Maybe a full fledged string theory would itself tell us how to
pose the right questions ultimately leading beyond itself.

At this point, we should remember Bohr's discussion of measurement in
quantum mechanics which in its basic ideas is applicable to any physical
theory, including string theory. The starting point of the argument is that
it does not make any sense to try to discuss the question what ``physics is
really like''. We have to be content with investigating how physics which
physicists can know about is like. But irrespective of how ingenious and
inventive experimental physicists and high tech engineers will be in the
future, any experiment will ultimately have to include some measuring device
which - for the purpose of the experiment under consideration - can be
supposed to work according to the laws of classical mechanics. The finger
pointing to the value of some measured quantity on a scale has to work as a
classical device. If you would observe it in e.g. funny quantum
superpositions, you would simply not be able to communicate about the
results of the experiments with a colleague or to give a conference report
about these. So, ultimately higher precision of an experiment has to
transform into higher precision of a classical measuring device but, simply
because it is a classical device, higher precision means higher precision in
the space-time coordinates (e.g. of the finger on the scale). Of course, we
can use different scales . We can observe the same effect on a, in total,
smaller or larger scale but this is simply a rescaling. It does not prevent
from the conclusion that if there is a limit to the increase of space-time
precision, there is a limit to the precision of any kind of experiment which
human physicists can perform.

So, the conviction that string theory is universal, in the sense that it is
not itself the limiting case of yet another more fundamental theory to be
dicovered one day by experiments of some very high precision, does not
derive from lack of fantasy to imagine e.g. some new exotic forces of nature
but from some very fundamental limits.

\bigskip

Can we really gain universal knowledge, even in principle? Of course, in
practice string theory will not provide you with the possibility to
calculate your success in the stock markets in advance. Since it adheres to
the laws of quantum mechanics, string theory has even intrinsic bounds on
the knowledge one can gain from it. But accepting the limits of Heisenberg's
uncertainty relation and of practical calculations, the idea of having one
day the final physical theory in our hands causes admiration and unease at
the same time. And it is not the purely melancholic kind of unease (``the
story is finally over, then'') which arises, here, but an unease which is
fostered by one of the deepest insights the 20th century has provided us
with.

Since G\"{o}del presented his incompleteness theorems, we know that
universality can require a price to pay. Hilbert's program of building
mathematics on formal axiomatics and proving that the used axiom systems are
free of contradictions, then, would - if successful - also have given
mathematics a final form. Indeed, in spite of G\"{o}del's results,
mathematics has found a base in the 20th century in the form of axiomatic
set theory which has proved to be universal so far in the sense that any
presently known field of mathematics can - in principle (again, this is
often not the most useful view in mathematical practice) - be boiled down to
constructions in this theory. The price G\"{o}del's incompleteness theorems
require us to pay for this is the never ending richness of models (in the
sense of model theory of formal axiomatic systems) which axiomatic set
theory has (and which - just by G\"{o}del's results - no formal axiomatic
theory whatever can come to terms with). It is the feeling that the
G\"{o}del results tell us a lesson about universality which goes beyond
their concrete form as a result about formal axiomatic theories which causes
the deeper form of unease one experiences when talking about the final
physical theory.

Comparing to G\"{o}del's results in formal mathematics, we expect that the
final physical theory, the ``theory of everything'', comes along with an
inexhaustible plentitude of models (here, we use the term ``inexhaustible''
to stress that we do not simply mean infinitely many models but also that
they can not be described in a single closed form). This expectation is not
in contradiction to the limits on observations we discussed above. Our
discussion of physical measurements above applies only to local properties.
Strictly speaking, the conclusion we can draw from it is that for any given
physical system (of limited volume) there is a final description upon which
we can not improve any more by increasing the precision of measurements. The
possibility that different local physical systems are described by different
models of the ``theory of everything'' and that there is an inexhaustible
plentitude of such models can not be excluded a priori. We will discuss this
point in more detail at the end of the next section.

Some warnings aiming at possible misunderstandings are in order, here. Since
we required that the models of the final theory can not be described in a
single closed form, we can not identify these with the usual initial
conditions of a physical theory. Also, we can not identify the models with
the different possibilities for Calabi-Yau compactification which arise in
string theory. The full background independent formulation of string theory
one hopes to discover should live on the full moduli space and therefore is
not supposed to be parametrized by different Calabi-Yau manifolds. How
should the different models arise, then? We will meet a natural candidate
for the price one has to pay for the universality of the ``theory of
everything'' in the next section.

\bigskip

\section{Is string theory quantum computational?}

Recent research has lead to the following two - partly conjectural -
scenarios:

\begin{itemize}
\item  The tangential structure of the moduli space (to be more precise, the
so called extended moduli space) of string theory at a Calabi-Yau manifold $%
W $ is described by the total Hochschild complex of the sheaf of holomorphic
functions on $W$ (\cite{Kon 1994}, \cite{Wit}).

\item  The possible deformation quantizations of a smooth finite dimensional
manifold $M$ constitute an infinite dimensional manifold $\mathcal{D}M$ and
the Grothendieck-Teichm\"{u}ller group acts transitively on $\mathcal{D}M$ (%
\cite{Kon 1999}).
\end{itemize}

\bigskip

At first sight, the two points may appear as completely unrelated. But the
second one is deeply tied to the Deligne conjecture (a proof of which is
announced in \cite{Kon 1999}) which states that the operad of chains of the
little discs operad naturally acts on the total Hochschild complex of 
\textit{any} associative algebra (and using a deep result of cohomology
theory, the so called Cohomology Comparison Theorem of \cite{GeSch},
therefore also on the total Hochschild complex of any sheaf of associative
algebras). Now, the automorphism group of the operad of chains of the little
discs operad is - at least in considerations up to homotopy - the
Grothendieck-Teichm\"{u}ller group. The generality of the Deligne conjecture
suggests, then, that the Grothendieck-Teichm\"{u}ller group might also
describe a fundamental symmetry of the (extended) mduli space of string
theory. We present a more detailed discussion of these issues elsewhere and
for our present purpose restrict to the remark that the idea that the
Grothendieck-Teichm\"{u}ller group could represent some fundamental symmetry
of string theory gets further support from the fact that the Kontsevich
formula for the deformation quantization of a finite dimensional Poisson
manifold $M$ can be derived from a two dimensional conformal field theory
with target space $M$ (see \cite{CF}, \cite{Kon 1997}). The weights of the
Feynman graphs of this two dimensional quantum field theory belong to a
subalgebra $P_{\Bbb{Z},Tate}$ of the algebra $P$ of periods (the notation
indicating that it is the algebra of periods of certain mixed Tate motives).
Again, it is conjectured (\cite{Kon 1999}) that - roughly speaking - the
symmetries of $P_{\Bbb{Z},Tate}$ are described by the
Grothendieck-Teichm\"{u}ller group (much the same way as the symmetries of
the algebraic numbers $\overline{\Bbb{Q}}$ are described by the absolute
Galois group $Gal\left( \overline{\Bbb{Q}}/\Bbb{Q}\right) $).

The reader who is not acquainted with part of the technical notions used
above should at least keep in mind the following two points for the
discussion in the rest of this section:

\begin{itemize}
\item  There is an algebra $P_{\Bbb{Z},Tate}\subseteq \Bbb{C}$ over $\Bbb{Q}$
which is conjectured to determine the Grothendieck-Teichm\"{u}ller group
which in turn might be a fundamental symmetry of string theory.

\item  The elements of $P_{\Bbb{Z},Tate}$ appear (as the weights of Feynman
graphs) in the deformation quantization of \textit{any} finite dimensional
Poisson manifold.
\end{itemize}

\bigskip

Let us speculatively suppose for the rest of this paper that the
Grothendieck-Teichm\"{u}ller group is a fundamental symmetry of string
theory. We could then - in a still more optimistic vein - hope that - as in
the case of other physical theories, e.g. general relativity - the symmetry
is linked to a basic principle which largely determines the theory. Is there
a candidate for a physical principle which determines the algebra $P_{\Bbb{Z}%
,Tate}$?

Let us try to give an answer by posing a seemingly quite different question:
What is the physically natural choice for an algebra of numbers? Since a
computer is a physical device, certainly all numbers which have a
realization on a computer have to be considered as physical. The rationals $%
\Bbb{Q}$ can be realized on computers and since the operations of arithmetic
have a realization, it is clear that the set of ``physical numbers'' has to
be an algebra over $\Bbb{Q}$. For a classical Turing machine (the model of a
universal computer), $\Bbb{Q}$ is already the maximal set of numbers we can
obtain. Even a simple non rational algebraic number like $\sqrt{2}$ has no
intrinsic realization on a Turing machine. Since we know that nature is not
classical but quantum mechanical, we should actually not consider classical
Turing machines but quantum computers. The rationals can, of course, be
realized there, too, and at first sight it seems that this is the final
answer, again. Concerning the question of principal computability, quantum
computers are equivalent to classical Turing machines, they only differ from
them with respect to computational complexity of problems (see \cite{Deu}).
But the Kontsevich quantization formula shows that there is a set of numbers
- which is precisely given by the algebra $P_{\Bbb{Z},Tate}$ - which is 
\textit{intrinsically} defined for \textit{any} deformation quantization of
a Poisson manifold (finite dimensional). So, if we assume a quantum computer
to be described by deformation quantization of some classical model, for a
quantum computer $P_{\Bbb{Z},Tate}$ should be the algebra of numbers which
are physically defined (Of course, assuming that there is a model for
quantum computation which can be described in terms of deformation
quantization is a true - i.e. non trivial - assumption which we will make in
the sequel. It is closely related to assuming that quantum computation can
be described by a low dimensional quantum field theory, cf. the work of \cite
{FLW}). Why is there no contradiction to the results of \cite{Deu}, here?
The answer is that \textit{in} the quantum computer only the rationals can
naturally be realized, as in the case of the classical Turing machine. The
way the periods in $P_{\Bbb{Z},Tate}$ arise in the process of quantization -
as weights of Feynman graphs - shows that they should be viewed as being
linked to a kind of scattering theory done on the system under
consideration. Scattering theory means testing the long time behaviour of a
system under possible input data. So, we expect that the numbers in $P_{\Bbb{%
Z},Tate}$ become only observable as intrinsic numbers of a quantum computer
when one tests the long term behaviour of the device with all possible kinds
of input data. Then they should emerge from the statistics of these tests.
Formulated in a down to earth way, one could say that the numbers in $P_{%
\Bbb{Z},Tate}$ should emerge as statistical coefficients from quantum
software testing (there are no intrinsic coefficients of this kind in the
classical case). In the language of formal logic these coefficients belong
to the meta level (since they refer to testing the device from the outside)
and not to the system itself. This is the reason why there is no
contradiction to \cite{Deu}. But testing a device is in every respect an
allowed physical procedure, too. So, a quantum computer should give the much
larger algebra $P_{\Bbb{Z},Tate}$ as the algebra of physically realized
numbers.

In conclusion, let us suggest the following physical principle for
determining the algebra $P_{\Bbb{Z},Tate}$:

\begin{enumerate}
\item  \textit{All} physical systems should be amenable to a \textit{real}
time simulation on a quantum computer and quantum computers should be
described as physical systems by deformation quantization of classical
Turing machines.

\item  The observable quantities of the world should be those which can be
determined by observation of quantum computers on quantum computers.
\end{enumerate}

\bigskip

The first part of this principle is quite close to a quantum version of the
Church-Turing hypothesis as discussed in \cite{Deu}. The real time
simulation of quantum systems on a quantum computer is possible for systems
with a finite dimensional quantum state space (i.e. for spin systems). The
Bekenstein bound on the entropy of systems of finite volume suggests that on
a fundamental level local physics should always be describable by systems
with finite dimensional quantum state spaces. There are indications that
this is true for string theory (see \cite{Sus}). We added the requirement
that a quantum computer is described by deformation quantization of a
classical Turing machine because this is, of course, decisive for our
arguments.

Part (1) of the principle means that we can in our mind replace the whole
world by a world consisting of quantum computers, only, without changing the
physical properties. The assumption of part (2) is quite natural, then. It
means that we should take this world of quantum computers at face value and
should consider only what quantum computers can determine internally and
externally (about each other) as physically observable.

Part (1) leads - as discussed above - to the algebra $P_{\Bbb{Z},Tate}$.
What is the effect of part (2)? We discussed the testing of quantum software
as if done by a classical agent. Part (2) would require to do the testing by
a quantum computer, too. But then the classical weights would have to be
replaced by operators themselves and, in effect, we would expect an algebra
of quantum periods and a $q$-deformation of the Grothendieck-Teichm\"{u}ller
group (instead of $P_{\Bbb{Z},Tate}$ and the classical
Grothendieck-Teichm\"{u}ller group) to arise as the fundamental symmetry
objects. The (extended) moduli space of string theory, we discussed at the
beginning of this section, is a space of classical backgrounds for string
theory (this is like considering the Ricci flat metrics as backgrounds for
perturbation expansions in general relativity). The full quantum theory will
have to include non classical backgrounds, too. We are not going more deeply
into this question, here, but it will be discussed in a forthcoming separate
publication where we will give technical arguments why one should, indeed,
expect the inclusion of non classical backgrounds to lead to quantum
deformations of $P_{\Bbb{Z},Tate}$ and the Grothendieck-Teichm\"{u}ller
group.

But part (2) of the suggested principle has even more radical consequences.
We can, in principle, imagine a quantum computer testing a quantum computer,
at the same time being tested itself by yet another quantum computer, and
even an iteration of this process. We would therefore have to deal with
iterative deformation quantization of the mathematical structures resulting
in each step. On the side of string theory this would mean that full
quantized string theory would itself be only one step on a ladder of
infinite quantization, an idea which has been uttered in the literature some
years ago (\cite{Gre}).

\bigskip

At this point we come back to the question of universality of string theory,
discussed in the previous section. The above principle leads us to the
conclusion that string theory is universal in the sense that the classical
theory we have to start from is universal (represented on the mathematical
side by the algebra $P_{\Bbb{Z},Tate}$ and the Grothendieck-Teichm\"{u}ller
group). On the other hand, there is the ladder of quantizations (on the
mathematical side represented by iterated quantum deformations of classical
mathematical structures). This ladder is comparable to the plentitude of
models of a formal axiomatic theory since it means that there is a whole
hierarchy of different ``quantum realizations'' of the universal classical
theory. This seems to be the price one has to pay for universality
(comparable to the price formal axiomatics has to pay as a consequence of
G\"{o}del's incompleteness results).

\begin{remark}
One could try to find a loop hole in our arguments in the following way:
When we argue that the observation of a quantum computer by a quantum
computer could be observed by a third one and that this should start a
ladder of quantizations, one could argue that one should actually describe
the first two computers as a single quantum system observed by the third
one. But this is only true if we neglect interactions or if interactions are
described approximately in a way (like by a Coulomb law) which still allows
for the application of the rules of simple quantum mechanics. We know that
on a more fundamental level interactions have to be described by quantum
field theories and then a ladder of quantizations arises (as we all know
from second quantization). One should keep in mind that the ``testing of
quantum software'' necessarily introduces a decisive interaction - without
which the full algebra $P_{\Bbb{Z},Tate}$ would not be observable - between
the devices. A short but intriguing discussion how the introduction of
interactions starts the ladder of quantizations can be found in the
introductory chapter of the first volume of \cite{GSW}.
\end{remark}

\begin{remark}
We mentioned an iterative application of $q$-deformation, above. For a
concrete example - a quantum deformation of quantum groups - the reader may
consult \cite{GS}.
\end{remark}

When discussing physical measurements in the previous section, we remarked
already that the different models - i.e. the different levels of
quantization - of the ``theory of everything'' can only be realized in terms
of different local physical systems. Observing a higher level of
quantization means that we have to have the possibility to observe a tower
of ever smaller (quantum) effects on a system. Let us recall once again that
the limit to this we found in the previous section does apply only to every
single system. It does not exclude the possibility that we can observe a
level of quantization which is non observable for one system by going to
another one. A system which allows for the observation of a larger part of
the hierarchy of quantizations has to be capable to provide more
information. Since this information can not be gained - as we have seen in
the previous section - by increasing the accuracy of position measurements,
there is only one possibility to provide this information: The system has to
be more complex. We therefore conclude that if the hierarchy of
quantizations is observable it has to correspond to a hierarchy of
complexity. The different models of the universal theory should correspond
to different levels of complexity of physical systems.

The above principle and the work of \cite{Gre} suggest - if one asks for the
observability of the ladder of quantizations - the following view on string
theory: The universality of the classical theory seems to reflect the fact
that it is a theory of all the fundamental forces and particles (and thus
the constituents of all kinds of matter), i.e. a ``theory of everything''.
But it does not seem to be a reductionist theory in the strict sense since
there seems to be a hierarchy of complexity corresponding to different
levels of quantization. For each level of complexity there would be a limit
on the information we can obtain from the system, i.e. we could observe the
level of quantization up to an $n$-th one but the $k$-th level for $k\geq n$
would be physically unobservable. One suspects that on the level of
elementary particle physics it is just the usual quantization of a physical
theory which is observable with all the rest of the ladder being non
observable.

\begin{remark}
We can not circumvent the levels of complexity by arguing that the system
just consists of elementary constituents. It is not a priori clear that we
can completely break apart the system in this way. In a theory involving a
ladder of quantizations there is no a priori rule which tells us that we
have to apply the same level of quantization to the composed system as to
the constituents. We expect that when a system becomes complex enough to do
an amount of ``quantum software testing'' which makes the algebra $P_{\Bbb{Z}%
,Tate}$ observable to it, the critical point for the next level of
complexity (or quantization) is reached.
\end{remark}

Concluding this section, we mention once again that interactions (the
``quantum software testing'') are inextricably linked to each passage to the
next higher level of quantization. In a theory following the suggested
principle, a ``cat'', belonging to a level of complexity which is at least
one step higher than the one of elementary particles and atoms, could
therefore never become a ``Schr\"{o}dinger's cat'' by an experiment
triggered by radioactive decay (following the rules of quantum mechanics, it
is not decisive for the ``cat'' actually to do ``quantum software testing''
on atoms but what matters is only the possibility that it is, by information
capacity, able to do it).

\bigskip

\section{Conclusion}

We have seen that there are strong arguments that string theory should have
the property of being the final physical theory but there may be a price for
this universality to pay in the form of a ladder of quantizations
(corresponding to a hierarchy of complexity). A possible fundamental
symmetry of string theory and the ladder of quantizations may both be
consequences of an underlying principle with a natural quantum computational
formulation. A paper dealing with some of the more technical aspects,
mentioned only shortly, here, will appear separately.

If the picture presented in this paper has anything to do with the true
nature of string theory, string theory would be both, an end and a
beginning. Since the days of antiquity human scientists and philosophers
have asked questions about space, time, and the ultimate constituents of
matter. String theory would be the final word about this, the end of a long
journey of investigation. But at the same time, it would be a new beginning,
the starting point of a physics asking questions about a quantum hierarchy
of complexity and about properties of quantum computation.

\appendix 

We have formulated our suggested principle in the language of quantum
computers because with quantum mechanics having become one of the standard
parts of modern mathematical physics, this should be an especially
intelligible form. But beyond the questions discussed in this article, the
work of Kontsevich (in \cite{Kon 1997}, \cite{Kon 1999}) could - in an
optimistic vein, once again - turn out to be the starting point for a new
understanding of quantization, itself. The decisive question is if the
universal infinite dimensional manifold $\mathcal{D}M$ of deformation
quantizations of a finite dimensional manifold $M$, as discussed in \cite
{Kon 1999}, can be determined by knowing that it is a principal homogenous
space of the Grothendieck-Teichm\"{u}ller group plus some natural
principles. Put into different words: The question is if the
Grothendieck-Teichm\"{u}ller group (and therefore - under the assumption of
a conjecture of \cite{Kon 1999} - the algebra $P_{\Bbb{Z},Tate}$) does -
modulo some natural principles - already determine what deformation
quantization is. If the answer to this question would be in the affirmative,
this would mean that quantization itself would be largely determined by two
simple yet deep principles. In the rest of this appendix we will explain
this idea in more detail, assuming speculatively, again, that the answer to
the above questions is indeed in the affirmative (i.e. the important step
remaining would be to determine the algebra $P_{\Bbb{Z},Tate}$, again).

We state the first of the two principles as:

\bigskip

\textbf{Principle 1:}

The physically relevant information of a state space of a physical theory is
fully contained in a suitable algebra of real valued functions (observables)
on state space and the deformation theory of this algebra.

\bigskip

The first part of this principle - physically relevant information is
contained in the algebra of observables on state space - is definitely true
for all the well established theories in physics. The second part - the
inclusion of the deformation theory of the algebra - can, of course, be
justified on abstract grounds: Why should a highly idealized mathematical
structure - like some algebra - be \textit{exactly} applicable to physics?
One would expect that algebras which are in some sense close to the original
one should also be of relevance, then. But this means precisely the
consideration of deformation theory of algebras. In string theory we indeed
consider the moduli space of two dimensional conformal quantum field
theories as the space of physically allowed classical backgrounds of the
theory. So, this principle is just a destilation of the structural core we
anyway accept for physical theories.

Let us come to the second principle, now:

\bigskip

\textbf{Principle 2}

The physically fundamental description of a Turing machine should have an
inherent definition for any numerical quantity which is of physical
relevance and can be explicitly defined (be it in arithmetic or geometric
terms).

\bigskip

Remember that the concept of a Turing machine, though a purely mathematical
concept on first sight, has a physical ingredient. In its classical form the
concept implicitly assumes that the machine is working according to the
rules of classical mechanics (as was pointed out by Deutsch in \cite{Deu}).
But - as we know definitely to be true nowdays from the existence of quantum
computers (though still very small in terms of bits, they exist) - this,
indeed, need not be the case and there are different, physically more
fundamental, concepts of a Turing machine than the classical one. Principle
2 assumes that on a fundamental level everything is accessible to (the
corresponding notion of) digital computers, there are no phenomena which are
fundamentally of an ``analog'' nature (to stress it once again, the
principle does in no way imply that this should hold for classical Turing
machines).

We can immediately see that a classical Turing machine does not satisfy
Principle 2 because only the rational numbers $\Bbb{Q}$ are inherently
defined quantities for a classical Turing machine. But as the Pythagoreans
observed already, there are other numbers like $\sqrt{2}$ which can equally
well be explicitly defined (e.g. as representing the diagonal in a square of
unit length). It is not important that by the atomistic nature of matter no
true diagonals of true squares can be built. In all the physical theories
known so far such geometric constructs still appear on a more abstract level
(e.g. in the setting of state space). E.g. $\sqrt{2}$ can also be defined
via normalization of an equal probability superposition of two quantum
states. So, we do not escape these quantities by refering to atomism.

All algebraic numbers $\overline{\Bbb{Q}}$ have an explicit definition
(either one refers to the corresponding algebraic equation or to a
geometrical construction) but still this is not sufficient. Volumes of
explicitly defined geometrical figures are also quantities we should
include. Since - by the foregoing considerations - we should consider a
geometrical figure to be explicitly defined if its boundaries can be defined
in algebraic terms, this means that we should include integrals over
algebraic functions on algebraic domains. But if one puts this into precise
mathematical terms it means that periods should have an inherent definition
in fundamental computing devices.

We should remember at this point that we have to apply Principle 1, too.
Assuming the validity of the partly conjectural scenario of \cite{Kon 1999}
(which includes especially the Deligne conjecture), it seems that only the
periods belonging to the algebra $P_{\Bbb{Z},Tate}$ are linked to the
deformation theory of associative algebras. So, both principles taken
together do indeed suggest the algebra $P_{\Bbb{Z},Tate}$ as the algebra of
numbers which a physically fundamental Turing machine should have an
inherent definition of.

We are done then in suggesting a way how to define a quantum computer from
first principles without presupposing quantum mechanics from the start. We
could use this definition of a ``fundamental Turing machine'' now to go back
and reformulate the principle suggested above for string theory by replacing
the notion of a quantum computer by that of a ``fundamental Turing machine''
(if one wants to have an equivalent formulation of this principle which does
not presuppose any conventional knowledge of quantization).

What does such an approach to quantization mean? It means that one tries to
understand quantization itself as being largely determined by a
computability requirement. We try to pass from simply observing the fact
that nature is intelligible (``the unreasonable effectiveness of mathematics
in the natural sciences'' which nowdays can be refined to the statement of
an unreasonable effectiveness of computer simulation) to turning this
observation into a fundamental physical principle. Only detailed future
investigations of the mathematical structures arising can lead to a
judgement if such an approach has a chance to lead to correct physics.

\end{document}